# Stronger Cryptography For Every Device, Everywhere: A Side-Channel-Based Approach to Collecting Virtually Unlimited Entropy In Any CPU

[1]JV Roig
[1]Advanced Research & Consulting – Asia Pacific College
[1]jvroig@gmail.com

*Abstract*— Generating secure random numbers is a central problem in cryptography that needs a reliable source of enough computing entropy. Without enough entropy available - meaning no good source of secure random numbers - a device is susceptible to cryptographic protocol failures such as weak, factorable, or predictable keys, which lead to various security and privacy vulnerabilities. In this paper, the author presents a significant improvement: a reliable way for any CPU-powered device - from the small, simple CPUs in embedded devices, to larger, more complex CPUs in modern servers - to collect virtually unlimited entropy through side channel measurements of trivial CPU operations, making the generation of secure random numbers an easy, safe, and reliable operation.

## I. INTRODUCTION

### A. Generating Random Numbers for Privacy and Security

The ability to generate strong random numbers is essential to cryptography, and central to security and privacy in the IT world. For our encryption technologies to function as expected, we rely on cryptographically secure pseudo-random number generators (CSPRNG) to actually produce high-quality random numbers suitable for cryptography. (For the purposes of this paper, "secure random number" will be used as shorthand to refer to "high-quality random number suitable for cryptography"; this refers mostly to how the random number was derived, and not any physical or inherent characteristic of any particular number itself)

Having a computer system generate a secure random number is a difficult task due to the deterministic nature of computers. Although random-generating algorithms have long existed, such as linear congruential generators (LCG) or the Mersenne Twister, these are not CSPRNGs (i.e., not suitable for use in cryptography). Hardware-based random number generators (called TRNGs or True Random Number Generators) are a common solution, particularly for servers that require massive amounts of entropy. TRNGs typically measure quantum random properties such as nuclear decay, or classical random properties such as thermal noise or atmospheric noise.

Being essential to security, some TRNGs have already made their way into CPUs themselves. VIA C3, released in 2003, has a TRNG built-in marketed as the VIA Padlock RNG [7], [22]. Intel also baked-in a TRNG into their CPUs starting in 2011 with the release of the Ivy Bridge architecture. [15], [17]

### B. Appliances and Devices

While traditional servers may consider the problem of generating secure random numbers solved due to easy access to TRNGs (a point I dispute in the next section), the world's security infrastructure does not rest solely in the hands of these servers. Heninger et al [18], [19] found widespread factorable and duplicate TLS and SSH keys due to embedded devices suffering from "boot-time entropy hole". This reveals a problem, mostly an economic/financial one, that also has to be solved: cheap devices and appliances do not have integrated TRNGs in them, and the resulting lack of entropy has caused a failure in their cryptographic protocols, which ended up producing thousands of duplicates of keys for TLS certificates in the wild and vulnerable RSA and DSA keys.

Ideally, manufacturers or vendors should shell out the money to make sure their devices deal with entropy (or its shortage), equipping their devices or appliances with a TRNG somehow. In the real world, pragmatism tends to miss out, and if the added cost of including a TRNG does not make financial sense, manufacturers and vendors will continue selling vulnerable devices.

A software solution here would be superior, especially if one can be applied cost-free (such as Open Source software) and the implementation is also simple and straightforward (i.e., not a significant burden to their existing development team)

### C. The problem with hardware-based TRNGs

In late 2013, due to the Snowden revelations, the TRNGs that were integrated into the CPUs themselves – Intel's DRNG and VIA's Padlock RNG – have fallen into suspicion [11], [12].

Whether the NSA has truly backdoored these by compelling Intel / VIA is not the central problem. The real problem is that these implementations are effectively blackboxes and are impossible to audit, especially in a live environment (there's no way you can actually audit the hardware circuit without destroying your CPU). For something so essential and central to cryptography and our security and privacy, we should not be depending on blackboxes.

Hardware devices also eventually fail. Aside from possibly being open and auditable, an acceptable software solution is superior to TRNGs in this regard, since software does not go bad like hardware does. Hardware random number generators do come with safety and health checks, but this is not a total safeguard – TRNG failure can result in service interruption that lasts until the specific device is replaced, since the computer system relying on it may have no other source of secure random numbers.

## II. BACKGROUND

### A. Shortcomings of the Current State of Secure Random Number Generation

Being essential to cryptography and central to our security and privacy, the generation of secure random numbers was a topic that interested me. Over the Christmas holidays of 2017, I thought about this problem, looked at the existing alternatives, and decided that the current state is severely lacking and needs to be improved. Specifically:

1) Reliance on hardware-based TRNGs must end or be mitigated – hardware devices effectively act as blackboxes, are impractical to audit in general, and next to impossible to audit within a live environment. Due to their blackbox nature, they are a natural target for nation-states and their intelligence apparatus. Whether the NSA / China / Russia has or has not backdoored (through trickery or coercion) any TRNG is not the real problem – the problem is that they have been given a target to backdoor due to widespread reliance on hardware random number generators that are next to impossible to audit.

2) Current software solutions are lacking – software solutions either do not have enough theoretical backing, or are so complex that it makes them impractical to review and audit, and thus prone to suspicion; sometimes, both are true [2], [3], [4], [14]. There is also no solution currently available that is purposely designed as an architecture-and-platform-agnostic heuristic.

### B. Timing Variability – the Benchmarker's Bane

Variability of benchmark runtimes has long been a bane for hardware reviewers, testers, researchers, developers, and most other users that rely on benchmark performance for key decisions (for example, whether a particular code change has actually sped up or slowed down a particular function). When measured with enough precision, benchmark runtimes can vary wildly, and are generally irreproducible. This applies to CPU benchmarks, GPU benchmarks, benchmarks of other hardware (hard disks, SSDs, etc.), and especially to benchmarks that combine many of these components.

This was where I first started to imagine what would end up as the basis of the SideRand prototype – can CPU benchmark variability be used as the basis of entropy collection to generate secure random numbers? This presents itself as an interesting target for research and testing, since depending on variance of a benchmark runtime means using a side channel measurement, instead of the actual value of any mathematical operation which would be deterministic. If made to work, relying on a side channel measurement of a CPU would go a long way to solve the 2 problems mentioned in the previous section:

1) Reliance on hardware TRNG will be removed or mitigated, since everything that has a CPU – from large servers to small embedded devices or appliances – can potentially benefit.
2) This type of software solution, being based on a side channel, may survive cryptanalysis, since it does not depend on an algorithm that may produce cyclical output.

### C. CPU Variance

Assuming for now that the timing variability can be made to collect enough entropy to be suitable for cryptography, one obvious shortcoming that needs to be addressed is the "same-CPU" weakness. That is, if one specific CPU (say, an Intel i7-7700K) produces 100,000 different unique runtimes for a specific benchmark (with the variations being measured in nanoseconds), can an attacker produce the same 100,000 unique runtimes (thereby potentially making the proposed side channel-based RNG predictable) if she buys the exact same CPU and runs the exact benchmark? This means a potential fatal weakness would simply be: *"find out potential possible CPUs running in the target's data center, buy these and make a table of potential values"*

Fortunately, the answer here is: CPU performance varies, even between two CPUs of the exact model, family and stepping. Researchers from Lawrence Livermore National Laboratory, for example, published a 2017 paper detailing an empirical survey of the variation in performance and energy efficiency in their clusters of servers [5]. After characterizing the performance and energy efficiency of 4,000 CPUs, they found that no two processors had identical performance characteristics. This variation has not been improving (i.e., not becoming less pronounced) as CPUs become more modern; instead, from Sandy Bridge ($2^{nd}$ generation Intel Core architecture) to Ivy Bridge ($3^{rd}$ generation) to Broadwell ($5^{th}$ generation), the variation in performance has increased between processors of the same model, family and stepping. In a nutshell, since no two CPUs perform identically (given enough precision in measurement), relying on a side channel measurement based on benchmark runtime is not trivially exploitable by merely purchasing the same CPU model.

## III. DESIGN OF SIDERAND

### A. Fundamental Design Notes

I developed SideRand based on the variability of benchmarks. Its design is guided by the principle I call *auditability* – making the code as simple as possible to make auditing very easy. The evolution of SideRand prototypes, from the first attempt (designated as "mark 1" or "mk1"), until the final prototype version (currently, SideRand mk10), can be reviewed at the author's SideRand site: http://research.jvroig.com/siderand. (A full discussion of each step of the evolution, however, is outside the scope of this paper. Notes about the evolution can be found in the SideRand site as supplemental information.) The mk1 to mk10 prototypes were developed in Python 3. This included the essential part of entropy collection (the aggregation of timing information from trivial CPU operations), plus the part that uses the entropy collected to produce actual random bits (a generic hash operation, specifically SHA256).

Once the final Python 3 SideRand prototype proved itself in internal testing, I created a new prototype, this time using the C language to test the theory using a compiled language. This prototype was more widely tested than the Python 3 prototypes. The rest of the paper deals exclusively with the C version of the SideRand prototype and its results on a plethora of different CPU types. This prototype is meant only to test the entropy collection portion (particularly to show that even compiled executables will show enough benchmark variability compared to its interpreted counterpart), and so it only has the entropy collection part, and does not have the hashing component present in the Python 3 prototypes before it.

### B. SideRand C Prototype

Two versions of the SideRand prototype were done in C: one using the normal *clock()* timer from *time.h*, and one using the nanosecond-level timer *clock_gettime()*, also provided by *time.h*. The source code is published below, as Algorithm 1 and Algorithm 2.

**ALGORITHM 1:** SideRand C – normal *clock()* timer

```c
#include <stdio.h>
#include <time.h>
int main() {
    int i=0;
    int j=0;
    int samples = 256;
    int scale = 5000000;
    int val1 = 2585566630;
    int val2 = 576722363;
    int total = 0;
    double times[samples];

    for(i=0; i<samples; i++)
    {
        clock_t begin = clock();
        for(j=0; j<scale; j++)
        {
            total = val1 + val2;
        }
        clock_t end = clock();
        double time_spent = (double)(end - begin) / CLOCKS_PER_SEC;
        times[i] = time_spent;
        printf("%f\r\n", times[i]);
    }
    return 0;
}
```

Algorithm 1 shows the source code (<30 line, including headers and bracket lines) of the SideRand C entropy collection version using the normal *clock()* timer. Given my focus on *auditability*, you'll note how simple and straightforward the code is. It's just a loop of a trivial addition operation, which we time repeatedly with a chosen timer (in this specific case, the *clock()* function). We collect the timing information in an array. In total we are getting 256 samples, i.e., 256 timing values. The complete array, with its sequence of timing information, is our collected entropy.

For more modern Linux distributions (Ubuntu and Fedora, for example), *clock()* has microsecond-level precision. For Windows and older, more conservative, enterprise-type Linux distributions (RHEL, CentOS), this has only millisecond-level precision, and unsuitable for our needs. An improved prototype is shown in Algorithm 2, which uses available nanosecond timer, *clock_gettime()* instead.

**ALGORITHM 2:** SideRand C – nanosecond timer

```c
#include <stdio.h>
#include <time.h>

struct timespec timer_start(){
    struct timespec start_time;
    clock_gettime(CLOCK_PROCESS_CPUTIME_ID, &start_time);
    return start_time;
}

long timer_end(struct timespec start_time){
    struct timespec end_time;
    clock_gettime(CLOCK_PROCESS_CPUTIME_ID, &end_time);
    long t_nanos = (end_time.tv_sec - start_time.tv_sec) * (long)1e9 + (end_time.tv_nsec - start_time.tv_nsec);
    return t_nanos;
}

int main() {
    int i=0;
    int j=0;
    int samples = 256;
    int scale = 5000000;
    int val1 = 2585566630;
    int val2 = 576722363;
    int total = 0;
    long times[samples];

    struct timespec t_start;
    long time_spent = 0;

    for(i=0; i<samples; i++)
    {
        t_start = timer_start();
        for(j=0; j<scale; j++)
        {
```

```
        total = val1 + val2;
    }
    time_spent = timer_end(t_start);
    times[i] = time_spent;
    printf("%ld\r\n", times[i]);
  }
  return 0;
}
```

This prototype with the nanosecond-level timer is only trivially longer at 42 line, still including headers, blank lines and bracket lines. The algorithm is essentially unchanged – a trivial addition operation is looped several times, timed using the nanosecond-level timer. The timing info is aggregated into an array, which becomes our collected entropy.

## IV. TESTING AND ENTROPY ANALYSIS

### A. "The Entropy Source We Deserve, But Not the One We Need Right Now"

The output of any random number generator cannot be used directly as proof of its suitability as a source for secure random numbers. Fortunately, our cryptographic protocols also do not require that the source is true randomness. Instead, cryptographically secure random number generators merely need to be sufficiently unpredictable and have a ridiculously large key space (set of possible outputs) such that brute-force attacks are infeasible within the applicable threat-model. To paraphrase Gordon in "The Dark Knight (2008)", "*True randomness is the entropy source we deserve, but not the one we need right now.*"

The C prototype was deployed to several machines, and the output (timing information) was collected in CSV files. All tests were triggered with *niceness* set to -20 (i.e., highest priority). These files were then processed in order to find out how many unique timing values were collected, and the exact frequency distribution. Knowing these two things – number of unique values and their frequency - allows us to effectively model the minimum entropy in the system that we are able to collect.

### B. Key Space

Several different machines were used to run the SideRand C prototype. The raw results, as well as the tool itself for the reader's own testing or inspection, can be found in the SideRand website linked to in section III. In summary, the key space analysis shows that the runtimes measured can have thousands of unique values, which depend largely on CPU speed, available timer resolution, and the software stack. Using the nanosecond timer boosts unique values to over 100,000 unique values.

With thousands of potential values and with the raw random output (i.e., the value that will be hashed) being a chronological sequence of 256 of these values (i.e., the specific order they were collected in, not sorted in any way), that gives a ridiculous upper limit of ~1,000 ^ 256 (*clock()* timer) or ~100,000 ^ 256 (*clock_gettime()* timer). This is the potential upper bound. It's the stuff that the wild imaginations of a crypto-nerd (such as yours truly) are made of. However, this potential key space will be greatly affected by predictability – it doesn't matter if there is an unimaginably large set of possible values if, in practice, there are really only a handful of values that appear 99.999% of the time. To determine this, we need to analyze the frequency distribution of the runtime values.

### C. Frequency Distribution

Unique timing values were tabulated according to how often they appeared (frequency), and the % of time they appeared was calculated (number of appearances of a unique value, divided by the total number of timing values collected). All experimental results (raw and processed, in CSV and spreadsheet form) are available in the SideRand website linked to in section III as supplementary information.

In summary, for most modern x86-64 computers, even just the microsecond-level timer resulted in the most frequent (repeating) timing value to occur less than 1% of the time. Using the nanosecond-level timer resulted in this figure going down to less than 0.001%.

The worst case was found, predictably, in smaller, less-complex CPUs (particularly, in-order-execution architectures). SideRand C was tested in a Raspberry Pi 3 board (ARM Cortex A53 CPU), and an old Intel "Diamondville" Atom 330 CPU (circa 2008). For the Cortex A53, the microsecond timer resulted in the most frequent value appearing 14% of the time, compared to 17% of the time for the Diamondville Atom CPU. Using the nanosecond timer, the Cortex A53 had the most frequent value appearing 1.29% of the time, compared to only 0.02% of the time for the Diamondville Atom.

### D. Modeling System Entropy – Minimum Entropy Estimate Based on Most Frequent Value

Modeling the entropy accurately based on raw timing values and frequency distribution is a challenge that I'll reserve for more enthusiastic statisticians. For now, all we need is to model a conservative minimum entropy estimate – something reasonable that is sure to be lower than what the actual entropy would be. If even that worst-case estimate of the minimum entropy meets our standards, then we don't have to care anymore about what a better, more accurate entropy estimate is.

One way we could model the minimum computational entropy is to simply consider the most frequent value that appears. For example, let's imagine a data set that has a most frequent value (MFV) that appears 20% of the time. We are guaranteed at least 5 possible states of equal probability (*states = 100%/MFV*), since all other values can only be less than or equal to 20%. A minimum of 5 equally possible states means ~2.32 bits of entropy per measurement (*bits = log$_2$(states)*). Since we're stringing along 256 of these measurements, that gives us at least 594 bits of entropy for this theoretical system.

Table I. Some selected, representative results of the SideRand benchmarks, showing different classes of CPUs.

| CPU | Arch | Out-Of-Order Execution | OS | Timer Precision | MFV | Entropy (bits) | Avg Time (seconds) |
|---|---|---|---|---|---|---|---|
| Cortex A53 (RPi 3) | ARM | No | Raspbian | Nanosecond | 1.29154% | 1,606.34 | ~13.00 |
| Cortex A53 (RPi 3) | ARM | No | Raspbian | Microsecond | 14.22320% | 720.30 | ~13.00 |
| Intel Atom 330 | x86 | No | Ubuntu 16.04 | Nanosecond | 0.02388% | 3,070.23 | ~5.70 |
| Intel Atom 330 | x86 | No | Ubuntu 16.04 | Microsecond | 16.83393% | 656.80 | ~5.70 |
| AMD E350 | x86 | Yes | Ubuntu 17.04 | Nanosecond | 0.00036% | 4,624.72 | ~8.50 |
| AMD E350 | x86 | Yes | Ubuntu 17.04 | Microsecond | 0.12555% | 2,467.21 | ~8.50 |
| Pentium G3260 | x86 | Yes | CentOS 7.5 | Nanosecond | 0.06435% | 2,714.06 | ~2.75 |
| Pentium G3260 | x86 | Yes | CentOS 7.5 | Microsecond | 7.94430% | 935.41 | ~2.75 |
| i7 7700K | x86 | Yes | Fedora 25 | Nanosecond | 0.00049% | 4,511.94 | ~2.80 |
| i7 7700K | x86 | Yes | Fedora 25 | Microsecond | 0.15527% | 2,388.75 | ~2.80 |

For a cryptographically secure seed, the standard to meet is an entropy of at least 256 bits (approximately $1.15 \times 10^{77}$), from which can be derived an unlimited number of keys using deterministic cryptography [8], [9], [10]. 128 bits of entropy was suggested by the IETF in 2005 [13]. To be more conservative, this paper will consider 256 bits as the standard to shoot for.

Experimental data gathered, as described in sections IV-B and IV-C, result in complete overkill. While we need only 256 bits of entropy, most modern x86-64 CPU architectures end up collecting over a thousand bits of entropy in a single run of the SideRand C prototype, even with just the microsecond-level timer. Using the nanosecond-timer, these CPUs collect between 3 to 5 thousand bits of entropy in a single run. Examining the worst cases – the Intel Diamondville Atom and the ARM Cortex A53 CPUs, using only the microsecond-level timer – still gives us well in excess of 650 bits (Diamondville Atom) or 720 bits (Cortex A53) in a single run.

Table I shows a summary of representative results from these experiments. The "CPU" column identifies the specific CPU model. The "Arch" column identifies the architecture of the CPU (either ARM or x86). The "Out-Of-Order Execution" column marks whether the CPU is an out-of-order execution architecture ("Yes") or not. "OS" identifies the OS used in the experiment. The "Timer Precision" column indicates the precision of the timer used. The "MFV" column reports the Most Frequent Value % found – that is, the frequency (in %) of the value that repeated the most (e.g., a 2% MFV means the most frequent value accounted for 2% of the entire data set, and all other values have a <=2% frequency). The "Entropy (bits)" column shows the bits of entropy gathered, based on the MFV, as described earlier in this section. Finally, the "Avg Time (seconds)" column reports the average runtime of SideRand in the given platform.

## V. WHERE THE VARIANCE COMES FROM

It may be worthwhile at this point to mention why this runtime variance between benchmark runs exists.

Modern CPUs contain millions to billions of transistors – even the ARM Cortex-A9, released over ten years ago (2007), has an estimated 26 million transistors. These transistors that make up a CPU are not perfectly uniform (as truly nothing humans create ever are, when measured with enough precision), and transistor variability has long been something that CPU designers cope with, such as designing for the worst case (guard-banding). Aside from transistor variability itself, there are also systematic variability (caused by manufacturing, with CPU binning as a common coping strategy) and local variability (random dopant fluctuation) [21]. These factors make CPUs unique from each other, even those that come from the same wafer and binned as the exact same model, family and stepping. CPU enthusiasts, especially overclockers, refer to this as the "silicon lottery".

This variation between CPUs that are sold as identical is not getting better (smaller). As the LLNL team found in [5], this variance has only increased with more recent processors (e.g., Broadwell microarchitecture compared to Ivy Bridge microarchitecture). The reason for this is the improvement in dynamic frequency scaling features in most CPUs – whereas these "turbo" features in multi-core architectures used to be very blunt (a fixed frequency if only 1 core is operating, a slightly lower fixed frequency with an additional core operating, etc), current implementations from Intel, AMD and ARM-based processors offer turbo-like features with smarter capabilities that take into account estimated current consumption, estimated power consumption, and processor temperature (respectively, Intel Turbo Boost Technology 2.0, AMD SenseMI / Extended Frequency Range, ARM DVFS technology) [1], [6], [16].

These variances that stem from the transistor level also affect repeated runs of the same physical CPU. Execution of the same instruction and data won't take the exact same path each time – i.e., each run would not be using the exact same transistors (be they transistors in the CPU registers, cache, or execution units). The physical location of the data, for example, also results in runtime variance due to latency differences – one run, for example, may have the data on physically one end of the cache, whereas the other run may have the same data on the other end of the cache. These variations can be as small as nanoseconds or fractions of nanoseconds.

Aside from the practical, real-world consideration of needing a CPU and platform that offers reasonable timer resolution, the only way that this side channel-based secure random number generation heuristic will fail is when our technology is able to do two magical things:

1) Create perfectly uniform transistors.
2) Fabricate a complex processor that contains billions of these perfectly uniform transistors without introducing any manufacturing variations (i.e., a perfectly uniform, flawless, manufacturing process)

Additionally, CPUs would have to be designed so that performance is not dynamic (no more temperature, current, voltage, and workload-based dynamic frequency scaling).

Until we reach this level of technology, which does not seem to be on the horizon, and CPUs somehow revert back to having non-dynamic performance scaling features, the proposed side channel-based heuristic is likely to remain a good candidate for ubiquitous secure random number generation across all our CPU-powered devices.

## VI. POTENTIAL IMPACT, USE CASES AND LIMITATIONS

Primarily, I envision SideRand as an open and auditable way for operating system kernels or hypervisors to seed their RNG. Initial seeding during OS installation or first boot is a problem that still needs solving [10], which has led to the vulnerabilities that Heninger et al have discovered [19]. While SideRand is easily extendable to be a general purpose CSPRNG, I'm not very concerned about that right now; there are already great CSPRNGs around, we just need to solve the problem of initial seeding – creating that first 256-bits of entropy in order for all these CSPRNGs to do their job properly. Being auditable also means that the very core of our cryptographic security – initial seeding and random number generation – is strengthened against potential backdoor attempts by nation-states.

For servers, SideRand can serve as a replacement for TRNGs. This greatly improves auditability, as SideRand is easily auditable (in its current implementations, around 40 lines of code), whereas TRNGs are impossible to audit in live environments. This will also serve to improve reliability. Hardware eventually fails, even known good hardware, whereas known good software does not, and software is far easier to patch in live environments compared to swapping out hardware in case of eventually discovered defects or needed tuning.

One run of SideRand can replace or complement the entropy collector and random seeder in servers, devices and appliances that would otherwise have poor entropy, solving the problems encountered by Heninger et al [19]. A single run of SideRand can produce a key, from which an unlimited number of keys can then be derived using standard deterministic cryptography, such as using the SideRand-produced key as the nonce or initialization vector to a block cipher in counter mode (e.g. AES-256-CTR) [9].

Headless or embedded devices that previously suffer from "boot-time entropy-hole" can run SideRand during boot for a few seconds to produce the needed strong key, and from there generate secure random numbers using traditional cryptography, such as the aforementioned construction of a secure block cipher in counter mode.

Since target devices may range from large servers to small devices with micro-controllers like Arduino or RaspberryPi, tuning issues will matter. All experimental results shown in this paper are overkill – from 2.5x to 18x more than the 256 bits of entropy needed. This means tuning for SideRand to run faster without sacrificing security is feasible by making the SideRand algorithm collect less samples.

Platforms that have lower timer resolution than microseconds (milliseconds and higher) have not been tested. These platforms may not be able to collect enough entropy through the SideRand heuristic within a practical time. These sorts of platforms should be extremely rare though – even Arduino devices have microsecond-level timers.

## VII. CONCLUSION

I presented SideRand, a heuristic and prototype for generating secure random numbers based on the inherent variability of benchmark runtimes, with its worst-case entropy estimate shown to far exceed the required entropy in order to be considered cryptographically secure. Experimental data from various platforms (different CPU architectures, operating systems, and timer precisions) have shown that the heuristic works to provide strong cryptography to practically all types of devices, from small, embedded systems to large servers. Effectively, this closes the boot time entropy-hole found by Heninger et al [18], [19]. The simplicity of the entropy gathering code serves to deter nation-state actors from backdooring OS RNG seeding.

ACKNOWLEDGMENTS

Thanks to Peter Bright (Technology Editor at Ars Technica) for suggestions that made this research much better: Testing on in-order-execution architectures and maximizing process priority. Both items ensure most sources of jitter are controlled for, and provide a clearer picture of the reliability of this side channel as a strong randomness source.